\documentclass{amia}
\usepackage{lipsum} 
\usepackage{booktabs} 
\usepackage{multirow} 
\usepackage[caption=false]{subfig}
\begin{document}

\title{Vision Foundry: A System for Training Foundational Vision AI Models}

\author{Mahmut S. Gokmen, MS$^1$, Mitchell A. Klusty, BS$^1$, Evan W. Damron, BS$^1$, W. Vaiden Logan, BS$^1$, Aaron D. Mullen, MS$^1$, Caroline N. Leach, BS$^1$, Emily B. Collier, MS$^1$, Samuel E. Armstrong, MS$^1$, V. K. Cody Bumgardner, PhD$^1$ }

\institutes{
    $^1$ Center for Applied AI, University of Kentucky, Lexington, KY
}

\maketitle
\section*{Abstract}

\textit{Self-supervised learning (SSL) leverages vast unannotated medical datasets, yet steep technical barriers limit adoption by clinical researchers. We introduce \textbf{Vision Foundry}, a code-free, HIPAA-compliant platform that democratizes pretraining, adaptation, and deployment of foundational vision models. The system integrates the \textbf{DINO-MX} framework, abstracting distributed infrastructure complexities while implementing specialized strategies like Magnification-Aware Distillation (MAD) and Parameter-Efficient Fine-Tuning (PEFT). We validate the platform across domains, including neuropathology segmentation, lung cellularity estimation, and coronary calcium scoring. Our experiments demonstrate that models trained via Vision Foundry significantly outperform generic baselines in segmentation fidelity and regression accuracy, while exhibiting robust zero-shot generalization across imaging protocols. By bridging the gap between advanced representation learning and practical application, Vision Foundry enables domain experts to develop state-of-the-art clinical AI tools with minimal annotation overhead, shifting focus from engineering optimization to clinical discovery.}
\section{Introduction}

Recent advances in self-supervised learning and foundational vision models have reshaped how computer vision systems are developed for domain-specific applications \cite{ssl_review_1}. Unlike supervised methods that depend on large labeled datasets, SSL approaches such as DINO \cite{dinov2_paper, register}, SimCLR \cite{simCLR_4}, and MAE \cite{MAE_5} learn meaningful representations directly from unlabeled data. Foundational models trained with these techniques produce rich and transferable embeddings that support downstream tasks including classification, segmentation, and anomaly detection \cite{wild_6, training_ssl_7}. A key strength of these models is their ability to learn structured feature spaces that capture color distributions, textures, edges, and higher-level semantic patterns, enabling broad generalization across datasets and tasks \cite{ssl_understanding_8}.

These capabilities are especially relevant in medical imaging, where obtaining expert-labeled data is both expensive and time-consuming \cite{ssl_medical_9, label_efficient_10}. Meanwhile, hospitals and research centers already store large quantities of unlabeled WSIs and CT/MRI scans that remain underutilized \cite{eval_ssl_11}. SSL allows these datasets to be transformed into rich, domain-adapted representations, enabling downstream models to be trained with substantially fewer labeled samples and reducing reliance on extensive annotation efforts \cite{survey_medical_13}.

Tasks such as segmentation, classification, regression, or clustering rely on transforming model-derived representations into meaningful outputs. While supervised models attempt to learn these outputs directly from labeled data, foundational models instead prioritize learning the underlying structure of the dataset, resulting in richer and more transferable embeddings \cite{ssl_analysis_14}. Lightweight downstream heads can be trained to produce task-specific predictions using these embeddings, and both contrastive and non-contrastive SSL objectives further strengthen feature quality \cite{dive_ssl_15}. This shared-compute paradigm reduces labeling requirements, accelerates model development, and improves generalization in settings where conventional supervised models struggle to transfer across datasets or populations \cite{eval_ssl_11}.

Despite these advantages, the practical use of SSL frameworks remains challenging. Existing repositories often require substantial technical expertise, complex configuration workflows, and an understanding of distributed training procedures \cite{project_monai, eval_ssl_11}. Although several tools have been introduced to lower the barrier to SSL training, many still demand considerable coding effort, lack the flexibility needed for medical imaging workflows, and provide only limited support for modern SSL training strategies \cite{project_monai, Lightly_ssl}.
Downstream development also presents challenges, as designing appropriate head architectures and training pipelines frequently requires domain knowledge and iterative experimentation. This highlights the need for an integrated, accessible system that enables domain experts to pretrain, fine-tune, and deploy foundational models without extensive machine learning expertise \cite{challenges_1}.

To address these challenges, we introduce \textbf{Vision Foundry}, a unified platform that simplifies the training, adaptation, and deployment of foundational vision models for medical imaging applications. Vision Foundry integrates \textbf{DINO-MX \cite{DINO-MX}}, a modular SSL framework supporting large-scale distributed training, parameter-efficient fine-tuning, domain-specific augmentation, and multi-expert learning strategies. The platform provides a secure dataset manager, automated DGX cluster scheduling, and interfaces for training lightweight downstream heads, thereby removing many of the technical barriers associated with SSL workflows.

Our contributions are threefold. First, we present Vision Foundry as the first end-to-end system enabling medical imaging researchers to pretrain and adapt foundational models on their own datasets without requiring programming expertise or distributed systems knowledge. Second, we incorporate the DINO-MX framework, extending the DINO family with modular components for PEFT \cite{PEFT}, domain-specific augmentation, multi-expert learning, and cross-magnification training. Third, we demonstrate Vision Foundry’s adaptability across diverse imaging domains, including neuropathology WSIs, lung tissue cellularity estimation, and coronary calcium scoring, showing that models trained within the platform generalize effectively across modalities and tasks.

\section{Methods}
To address the complexities of medical AI development, Vision Foundry is engineered as a vertically integrated system that unifies data management, compute orchestration, and model training. In this section, we detail the technical implementation of the platform, beginning with the high-level architecture and the core \textbf{DINO-MX} framework. We subsequently describe the governance protocols that secure sensitive imaging data, the automated pipelines for distributed training and downstream adaptation, and finally, the experimental design used to validate the system's efficacy across histopathology and radiology domains.

\subsection*{Overall Architecture}

\begin{figure}[H]
 \centering
 \includegraphics[width=1.0\textwidth]{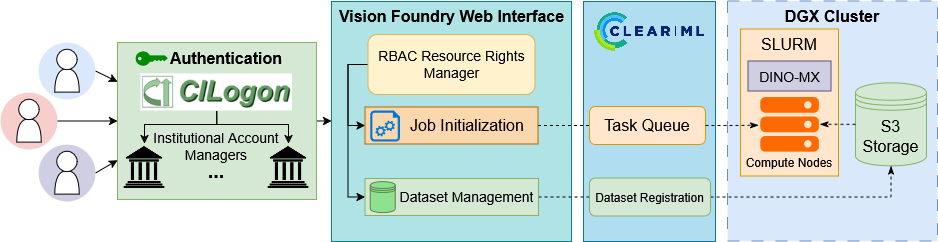}
\caption{\textit{Vision Foundry system overview. The platform unifies data management, distributed training orchestration, and model deployment within a secure infrastructure.}}
\label{fig:system_overview}
\end{figure}

The Vision Foundry platform is designed as a cohesive ecosystem that bridges the gap between raw medical imaging data and deployable artificial intelligence models. As illustrated in Figure~\ref{fig:system_overview}, the system architecture integrates a secure web interface, high-performance computing resources, and scalable object storage into a unified workflow. In this section, we detail the technical implementation of these core components: the \textbf{DINO-MX} framework, which serves as the computational engine for modular self-supervised learning; the \textbf{data governance} and storage infrastructure that ensures security and HIPAA compliance; the \textbf{training management} pipeline that orchestrates distributed workloads on the DGX cluster; and the mechanisms for developing task-specific \textbf{downstream heads} for clinical applications.

\subsection*{The DINO-MX Framework}

At the core of Vision Foundry lies \textbf{DINO-MX} (Modular \& Flexible), a specialized self-supervised learning framework designed to serve as the system's computational engine. While Vision Foundry handles data governance and job orchestration, DINO-MX executes the model training logic, abstracting the complexities of distributed computing and architecture definition. We designed DINO-MX to address the rigidity of existing research repositories, creating a flexible backbone that standardizes Vision Transformer (ViT) architectures for seamless compatibility with the Hugging Face ecosystem. This ensures that any model trained within Vision Foundry can be immediately deployed, shared, or fine-tuned using standard open-source tools without proprietary conversion scripts.

\begin{figure}[H]
  \centering
    \includegraphics[width=1.0\textwidth]{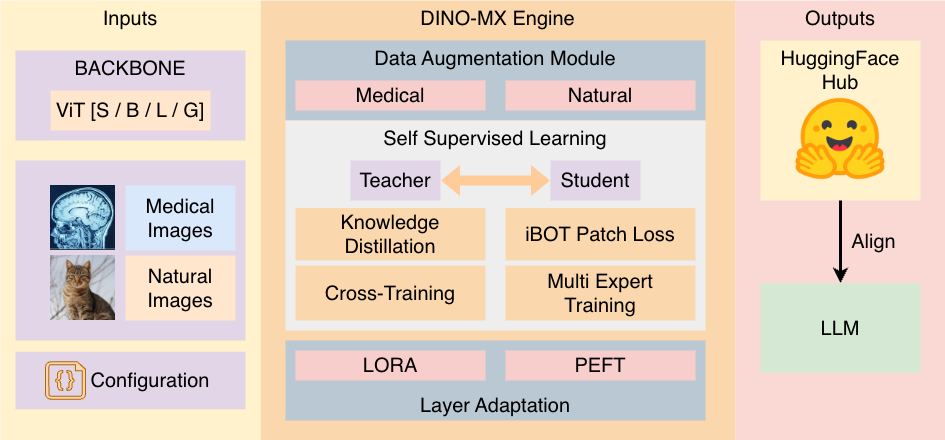}
 \caption{General representation of the DINO-MX framework}
 \label{fig:amia_figure}
\end{figure}
\textbf{Optimization for Compute Efficiency.} 
Medical imaging datasets often consist of gigapixel whole-slide images or high-resolution volumetric scans, demanding substantial computational resources. To make foundational model training feasible on standard institutional hardware, DINO-MX integrates advanced optimization techniques. It natively supports Fully Sharded Data Parallel (FSDP) and Distributed Data Parallel (DDP) strategies, enabling efficient scaling across multi-GPU nodes. Furthermore, the framework implements Parameter-Efficient Fine-Tuning (PEFT) \cite{PEFT} techniques, specifically Low-Rank Adaptation (LoRA) \cite{LoRA}. This allows Vision Foundry users to adapt massive pretrained backbones to new domains by training only a fraction of the parameters, significantly reducing memory overhead. Coupled with gradient checkpointing, these features allow the training of large-scale ViT architectures on resource-constrained environments.

\textbf{Domain-Specific Adaptation.} 
Unlike general-purpose computer vision libraries, DINO-MX is tailored for the unique modalities found in healthcare. The framework includes specialized data loaders and augmentation pipelines for medical formats. This includes native handling of single-channel inputs for CT and MRI modalities, as well as stain-invariance augmentations specifically designed for H\&E histopathology. Users can select these domain-specific configurations directly through the Vision Foundry interface, ensuring that the self-supervised objectives are aligned with the visual properties of the target medical data.

\textbf{Advanced Distillation and Multi-Expert Learning.} 
DINO-MX extends the standard DINO methodology with modular training strategies designed for complex medical distributions. It supports Multi-Teacher Distillation, allowing a student model to aggregate knowledge from multiple teacher checkpoints simultaneously to improve stability. Additionally, the framework implements a Multi-Expert Training strategy to handle severe class imbalance, a common issue in medical datasets. In this mode, the data is sharded across specialized "expert" models on separate GPUs, which learn independently via self-distillation before their knowledge is aggregated into a single generalized model via Exponential Moving Average (EMA) updates.

\textbf{Magnification-Aware Distillation (MAD).} 
To specifically address the multi-scale nature of digital pathology, DINO-MX introduces Magnification-Aware Distillation (MAD). Existing SSL methods typically treat different magnification levels as independent views, losing the hierarchical context inherent in WSIs. MAD modifies the standard teacher--student framework by feeding the teacher network with low-magnification global context (e.g., $10\times$) while the student network processes spatially aligned high-magnification details (e.g., $40\times$). By forcing the student to match the teacher's contextual representation, the model explicitly learns the semantic relationship between tissue architecture and cellular morphology. This capability, exposed through Vision Foundry, allows researchers to train models that maintain consistent representations across zoom levels without manual annotation.
\subsection*{Data Management and Governance}

While DINO-MX provides the computational engine for representation learning, the practical utility of medical AI is fundamentally constrained by data accessibility and security. To address this, Vision Foundry implements a centralized data lake architecture hosted on a locally maintained, HIPAA-compliant NVIDIA DGX cluster. The storage backend relies on Dell PowerScale running OneFS, which exposes an S3-compatible object storage interface. This architecture provides the high I/O throughput necessary to saturate GPUs during the training of large foundational models while ensuring the fault tolerance and encryption-at-rest required for sensitive medical data. To bridge the gap between strict institutional security protocols and research agility, the platform employs a federated identity model via OAuth 2.0 with a CILogon client, integrating directly with institutional Single Sign-On (SSO) services.

Once authenticated, user authorization is governed by a granular Role-Based Access Control (RBAC) system. Access to specific datasets, defined as rights-managed prefixes within the S3 bucket, is managed through a three-tiered permission structure (\textit{read}, \textit{write}, \textit{manage}) that enforces the principle of least privilege. To handle the transfer of massive medical imaging files efficiently, Vision Foundry utilizes S3 pre-signed URLs. When a user initiates an upload or download, the system generates a secure, time-limited URL that allows the browser to transfer data directly to the storage server. This mechanism bypasses the web application server, removing a critical bottleneck and ensuring that data transmission remains both fast and encrypted, thereby providing a robust, auditable foundation for collaborative research.

\subsection*{DGX Training Management}

Building upon this secure data infrastructure, Vision Foundry provides a streamlined interface for orchestration and execution of training workloads on the same 5-node, 40-GPU H100 DGX cluster. Users initiate jobs through an interactive web form that abstracts the complexity of the underlying DINO-MX framework. This interface allows researchers to configure essential parameters, such as the target dataset, distributed training strategy (DDP or FSDP), global batch size, and learning rates, without writing code. Furthermore, it exposes advanced configuration options for Low-Rank Adaptation (LoRA), including rank, scaling factors, and dropout rates, enabling precise customization for parameter-efficient fine-tuning tasks.

Once configured, the training workflow is managed by an automated orchestration pipeline designed for reproducibility and efficient resource utilization. The configuration is first sent to a ClearML task queue, where a monitoring service detects the request and submits it as a Slurm job as GPU resources become available. Each job launches a containerized DINO-MX instance that retrieves the specific dataset from S3 storage and executes the training logic. This containerization ensures that the training environment is consistent and isolated, preventing dependency conflicts between users.

Throughout the training lifecycle, the system provides real-time observability. The DINO-MX container streams live metrics, including global and local losses, teacher momentum, and GPU utilization, back to the Vision Foundry dashboard, allowing users to monitor convergence and detect anomalies instantly. Upon completion, model artifacts such as checkpoints and final weights are automatically pushed to the S3 server. Access to these artifacts is governed by the same RBAC system used for datasets, ensuring that the entire lifecycle of the model, from raw data to trained weights, is secure, auditable, and easily shareable among authorized collaborators.

\subsection*{Task-Specific Adaptation and Inference}

To translate the general representations learned by the foundational model into actionable clinical outputs, Vision Foundry provides a streamlined workflow for developing task-specific adaptation modules. Instead of retraining the entire architecture for each new clinical objective, the system uses the pretrained backbone as a frozen feature extractor and attaches lightweight projection layers to transform the resulting embeddings. This design enables researchers to build high-performance, domain-specific tools such as tissue segmentation models or disease classifiers with minimal computational cost and substantially reduced annotation requirements.

For spatially resolved tasks, the platform includes a dedicated interface for training pixel-level segmentation modules. Users supply annotated regions of interest in GeoJSON format, which the system converts into dense supervision targets. A specialized decoding layer is then trained to predict segmentation masks that align with these geometries. Once trained, these modules can operate in a human-in-the-loop setting by automatically producing candidate masks for unlabeled images, significantly accelerating dataset curation through label propagation.

In parallel, the system supports categorical decision-making tasks through the integrated \textit{CLASSify} module \cite{classify}. For applications such as phenotype identification or risk stratification, users provide a limited number of labeled examples to define target classes. Leveraging the rich feature space produced by the backbone, the system trains linear classifiers or shallow neural networks that efficiently predict class membership. Both segmentation and classification workflows can be deployed as large-scale inference jobs using the platform’s containerized orchestration pipeline, allowing researchers to run their specialized models on extensive unlabeled datasets and retrieve structured predictions directly from centralized storage.
\subsection*{Validation Strategy}

To demonstrate the versatility of Vision Foundry across diverse medical imaging domains and analysis tasks, we designed a comprehensive evaluation framework covering pixel-level segmentation, scalar regression, and volumetric risk stratification.

\textbf{Neuropathology WSI Segmentation.}
To validate segmentation performance on gigapixel pathology slides, we utilized a dataset of hematoxylin and eosin (H\&E) stained neuropathology whole-slide images (WSIs). The dataset was manually annotated by experts into six distinct regions: \textit{Gray Matter}, \textit{White Matter}, \textit{Leptomeninges}, \textit{Superficial Cortex}, \textit{Background}, and \textit{Exclude}. From these annotated regions, we randomly sampled a balanced training set of 50,000 tiles ($224\times224$ pixels) per class at both $10\times$ and $40\times$ magnifications, assigning labels based on the majority class of the tile.

For the experimental setup, we employed the \textit{IBI-CAAI/MAD-NP} \cite{IBI-CAAI_MAD-NP} foundational model, trained via DINO-MX, as the frozen feature extractor. A lightweight Multilayer Perceptron (MLP) segmentation head with a dropout rate of 0.1 was attached to the backbone. The combined dataset was partitioned using a randomized index permutation, with 15\% of data held out for validation. A separate, held-out test set comprising 103,926 tiles was reserved strictly for final evaluation. Training was executed for 60 epochs with a batch size of 64 (effective batch size of 32), a learning rate of $1\times10^{-4}$, and gradient accumulation steps of 4, utilizing early stopping with a patience of 30 epochs to prevent overfitting.

\textbf{Lung Tissue Cellular Density Estimation.} 
We further evaluated the system's regression capabilities by addressing a high-throughput computational pathology task: predicting total cell density from low-magnification tiles to bypass the computational cost of high-resolution processing. To generate automated ground truth, we processed 12 H\&E-stained lung tissue WSIs using Cellpose \cite{cellpose} at $10\times$ magnification. The resulting detection maps were spatially aggregated, summing counts from sixteen $10\times$ tiles, to create scalar regression targets for the corresponding $2.5\times$ tiles ($224 \times 224$ pixels).

To assess the impact of domain adaptation, we compared existing foundational models against a DINO-MX fine-tuned model pretrained on a curated dataset of 96 lung tissue WSIs (approx. 1.3 million tiles). A linear regression head was trained on the frozen $2.5\times$ embeddings to predict the aggregated cell counts. We employed a rigorous 4-fold cross-validation strategy at the slide level (training on 9 slides, inferencing on 3 per fold) to ensure complete separation between training and testing distributions.

\textbf{Coronary Calcium Risk Detection.} 
To assess the platform's utility in volumetric radiology, we developed a pipeline for Coronary Artery Calcium (CAC)\cite{CAC_2} assessment using the internal \textit{Heartlens dataset}. The primary challenge in this domain is the discrepancy between specialized cardiac scans (ECG-gated), which are synchronized with the heartbeat to freeze motion, and standard chest scans (non-gated), which suffer from significant motion blur. We utilized the \textit{CARD-ViT} backbone, trained via DINO-MX exclusively on 2,651 high-quality \textit{gated} CT scans. A lightweight segmentation head was then trained to localize calcified lesions at the pixel level.

The inference pipeline integrated a lesion-specific Agatston scoring module, which calculates the standard clinical risk score based on the area and density of detected calcifications. To evaluate the robustness of the features learned by Vision Foundry, we tested this model on the external Stanford Coronary Calcium dataset \cite{coca_dataset}. Critically, we evaluated the model on \textit{non-gated} scans without showing it any non-gated data during training. This experimental design serves to validate whether the self-supervised representations can bridge the significant domain gap between clean, synchronized cardiac images and noisy, standard chest scans without requiring site-specific retraining.
\section{Results} 
We evaluate the capabilities of the Vision Foundry platform through three distinct clinical case studies: multi-class segmentation of neuropathology whole-slide images, cellular density estimation in lung tissue, and coronary artery calcium scoring from CT scans. These experiments were designed to assess the quality of self-supervised representations across different modalities (histopathology and radiology) and task types (segmentation, regression, and classification). Furthermore, we examine the models' ability to generalize to unseen data distributions, specifically testing performance on external datasets with differing acquisition protocols. The following subsections detail the quantitative comparisons against general-purpose foundational baselines and analyze the clinical relevance of the generated predictions.

\textbf{Neuropathology WSI Segmentation}
The quantitative assessment of pixel-level segmentation, detailed in Table \ref{tab:combined_performance}, confirms that the representations learned via Vision Foundry are highly discriminative. The model demonstrated exceptional fidelity in delineating anatomical structures, achieving an F1-Score of 0.95 for \textit{White Matter} and 0.90 for \textit{Superficial Cortex}. Notably, the \textit{Background} class was identified with high precision (0.91), indicating effective suppression of slide artifacts.
\begin{figure}[H]
  \centering
  \includegraphics[width=1.0\textwidth]{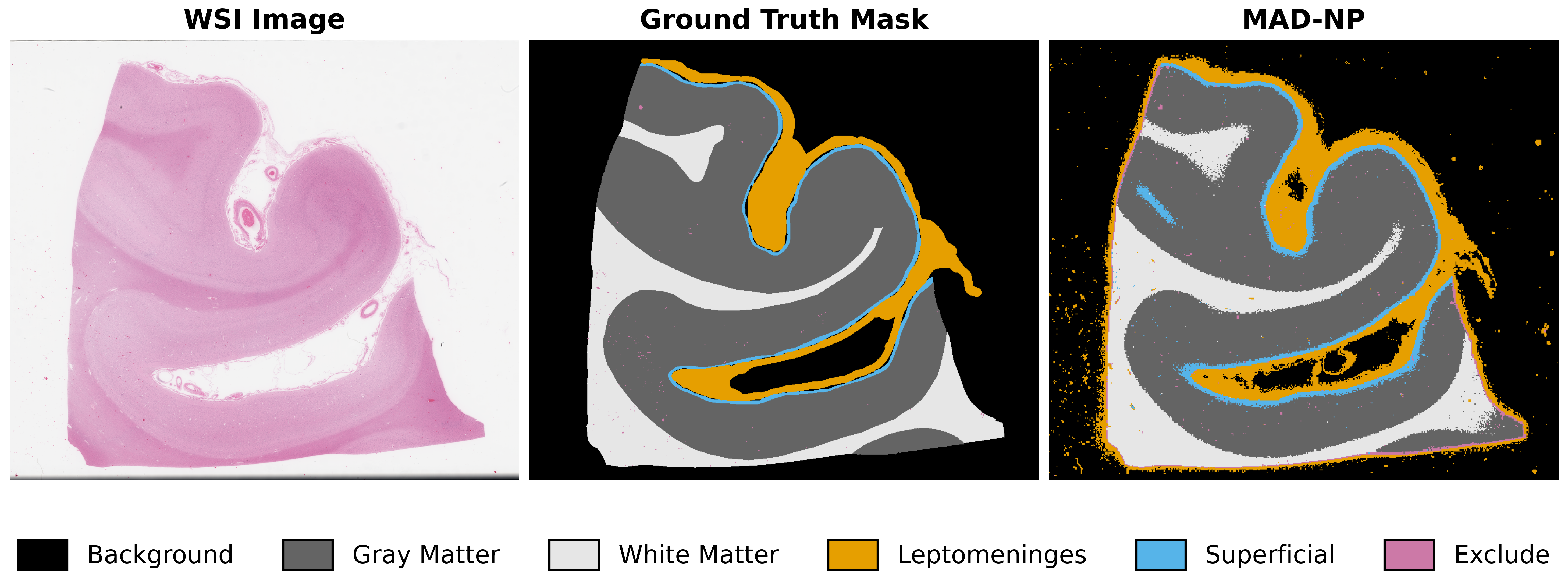}
  \caption{\textit{Qualitative segmentation results. Left: Original H\&E WSI. Center: Ground Truth annotations. Right: MAD-NP prediction showing precise delineation of cortical layers.}}
  \label{fig:neuropath_seg}
\end{figure}

Figure \ref{fig:neuropath_seg} provides a visual confirmation of these quantitative results. As illustrated, the MAD-NP prediction closely mirrors the ground truth annotations, successfully identifying distinct anatomical regions. The model demonstrates particular strength in delineating fine structures, such as the \textit{Leptomeninges} (red) and \textit{Superficial Cortex} (green), while maintaining sharp and accurate boundaries against the adjacent \textit{Gray Matter}. This visual evidence confirms that the learned embeddings are sufficiently discriminative to support high-fidelity tissue segmentation.

\begin{table}[H]
\centering
\caption{\textit{Evaluation of Neuropathology Segmentation. The top panel details the confusion matrix and per-class metrics for MAD-NP. The bottom panel compares MAD-NP's aggregate performance against state-of-the-art foundation models.}}
\label{tab:combined_performance}
\setlength{\tabcolsep}{2pt}
\small
\begin{tabular*}{\textwidth}{@{\extracolsep{\fill}}l|cccccc|ccc}
\toprule
\multicolumn{10}{c}{\textbf{A. MAD-NP: Confusion Matrix and Class-wise Performance}} \\
\midrule
& \multicolumn{6}{c|}{\textbf{Confusion Matrix (Predicted Counts)}} & \multicolumn{3}{c}{\textbf{Performance Metrics}} \\
\textbf{True Label} & \textbf{Back} & \textbf{Exc} & \textbf{Gray} & \textbf{Lepto} & \textbf{Superf} & \textbf{White} & \textbf{Prec} & \textbf{Rec} & \textbf{F1} \\
\midrule
\textbf{Background} & \textbf{13547} & 140 & 81 & 1310 & 11 & 3 & 0.91 & 0.90 & 0.90 \\
\textbf{Exclude} & 402 & \textbf{7588} & 2035 & 1128 & 926 & 658 & 0.75 & 0.60 & 0.66 \\
\textbf{Gray Matter} & 9 & 1522 & \textbf{22100} & 10 & 919 & 675 & 0.88 & 0.88 & 0.88 \\
\textbf{Leptomeninges} & 938 & 191 & 1 & \textbf{13140} & 67 & 12 & 0.84 & 0.92 & 0.87 \\
\textbf{Superficial C.} & 0 & 381 & 391 & 99 & \textbf{12759} & 20 & 0.87 & 0.93 & 0.90 \\
\textbf{White Matter} & 1 & 306 & 401 & 37 & 24 & \textbf{22094} & 0.94 & 0.97 & 0.95 \\
\bottomrule
\end{tabular*}
\vspace{0.3cm}
\begin{tabular*}{\textwidth}{@{\extracolsep{\fill}}lcccccc}
\toprule
\multicolumn{7}{c}{\textbf{B. Benchmark Comparison: Aggregate Metrics Across Models}} \\
\midrule
\textbf{Metric} & \textbf{MAD-NP} & \textbf{dinov2-giant} & \textbf{Prov-GigaPath} & \textbf{UNI2} & \textbf{UNI} & \textbf{Virchow2} \\
\midrule
\textbf{Avg. Linear F1} & \textbf{0.8601} & 0.8394 & 0.8570 & 0.8525 & 0.8526 & 0.8389 \\
\textbf{Avg. k-NN F1} & \textbf{0.8526} & 0.8315 & 0.8439 & 0.8412 & 0.8553 & 0.8201 \\
\textbf{Global AMI} $\uparrow$ & \textbf{0.6307} & 0.3816 & 0.4194 & 0.4291 & 0.2975 & 0.3369 \\
\textbf{Global DBI} $\downarrow$ & \textbf{1.6954} & 1.9780 & 2.1774 & 1.8774 & 1.9559 & 1.7680 \\
\bottomrule
\end{tabular*}
\end{table}
Comparative benchmarking (Table \ref{tab:combined_performance}, Panel B) further highlights the advantages of the magnification-aware training strategy. MAD-NP surpassed all general-purpose pathology models (Prov-GigaPath\cite{gigapath}, UNI\cite{uni}, Virchow2\cite{virchow2}) and the standard DINOv2-giant baseline across all aggregate metrics. Specifically, the high Global Adjusted Mutual Information (AMI) of 0.6307, markedly higher than the next best model, UNI2 (0.4291), implies that Vision Foundry produces a feature space where anatomically distinct regions form tighter, more separable clusters than competing approaches.

\begin{figure*}[t!] 
\centering
\subfloat[10x Cellpose Results]{\includegraphics[width=0.32\textwidth]{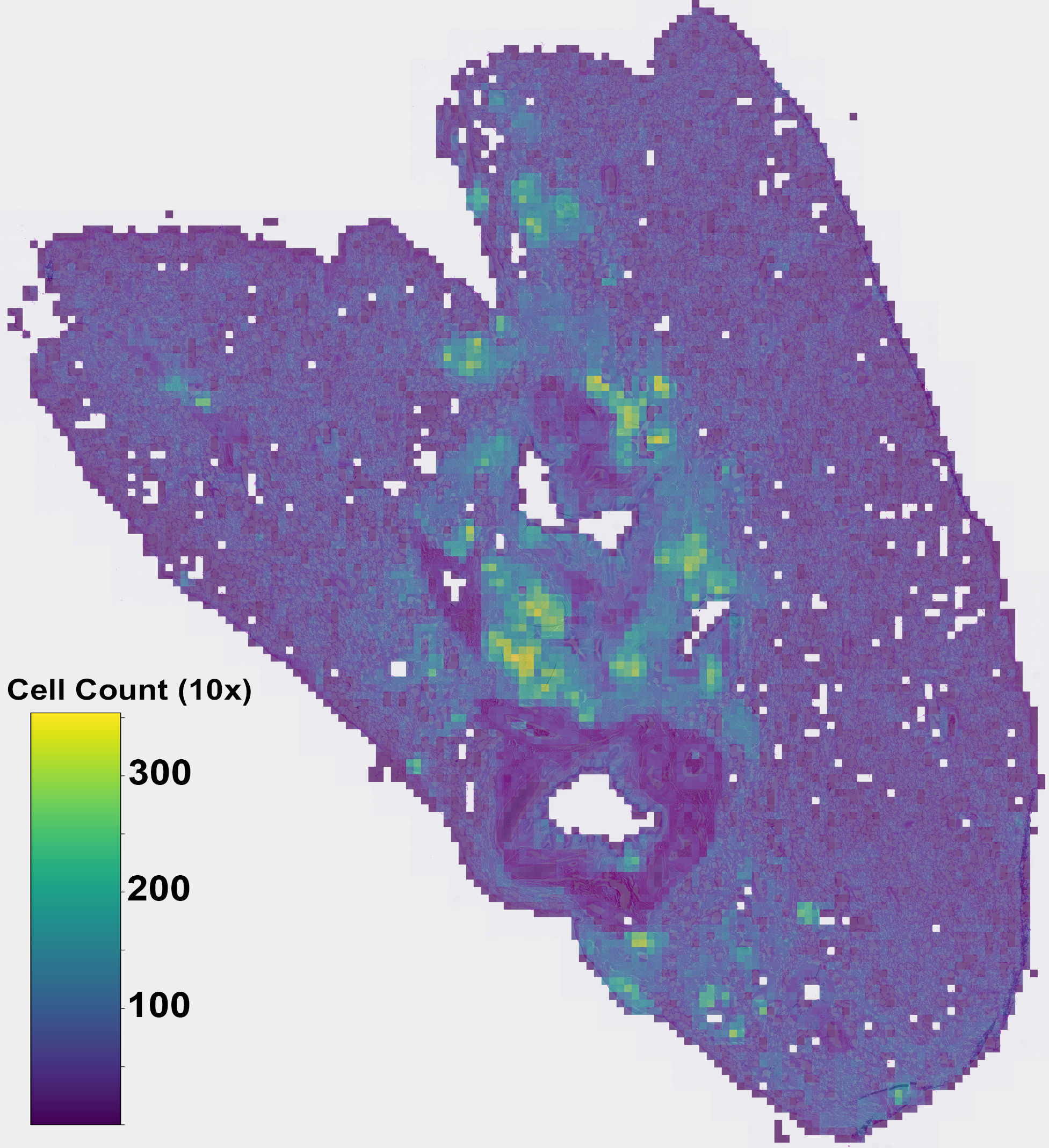}\label{fig:heatmap_10x_A}} \hfill
\subfloat[2.5x Aggregated Target]{\includegraphics[width=0.32\textwidth]{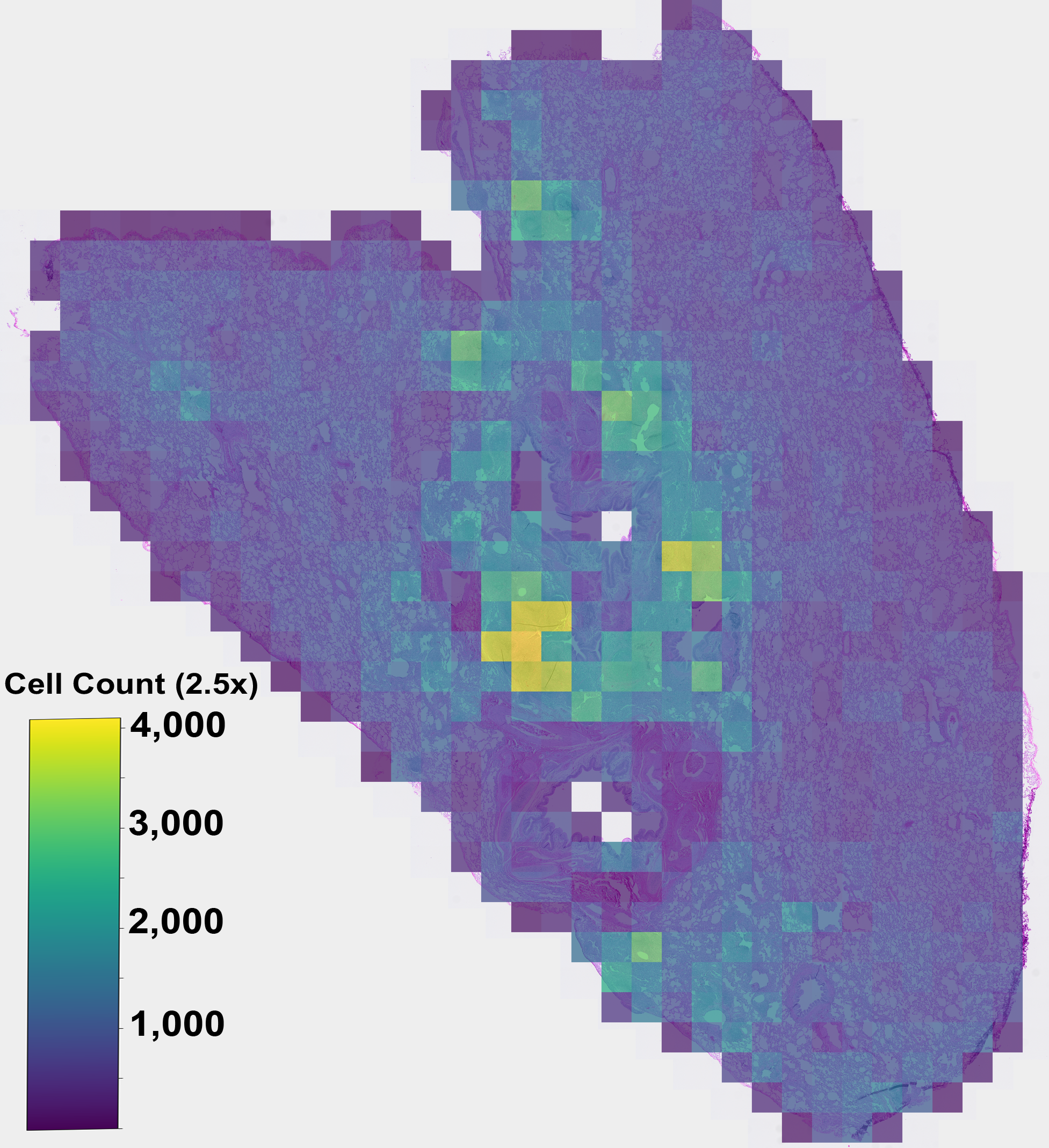}\label{fig:heatmap_agg}} \hfill
\subfloat[2.5x Vision Foundry Prediction]{\includegraphics[width=0.32\textwidth]{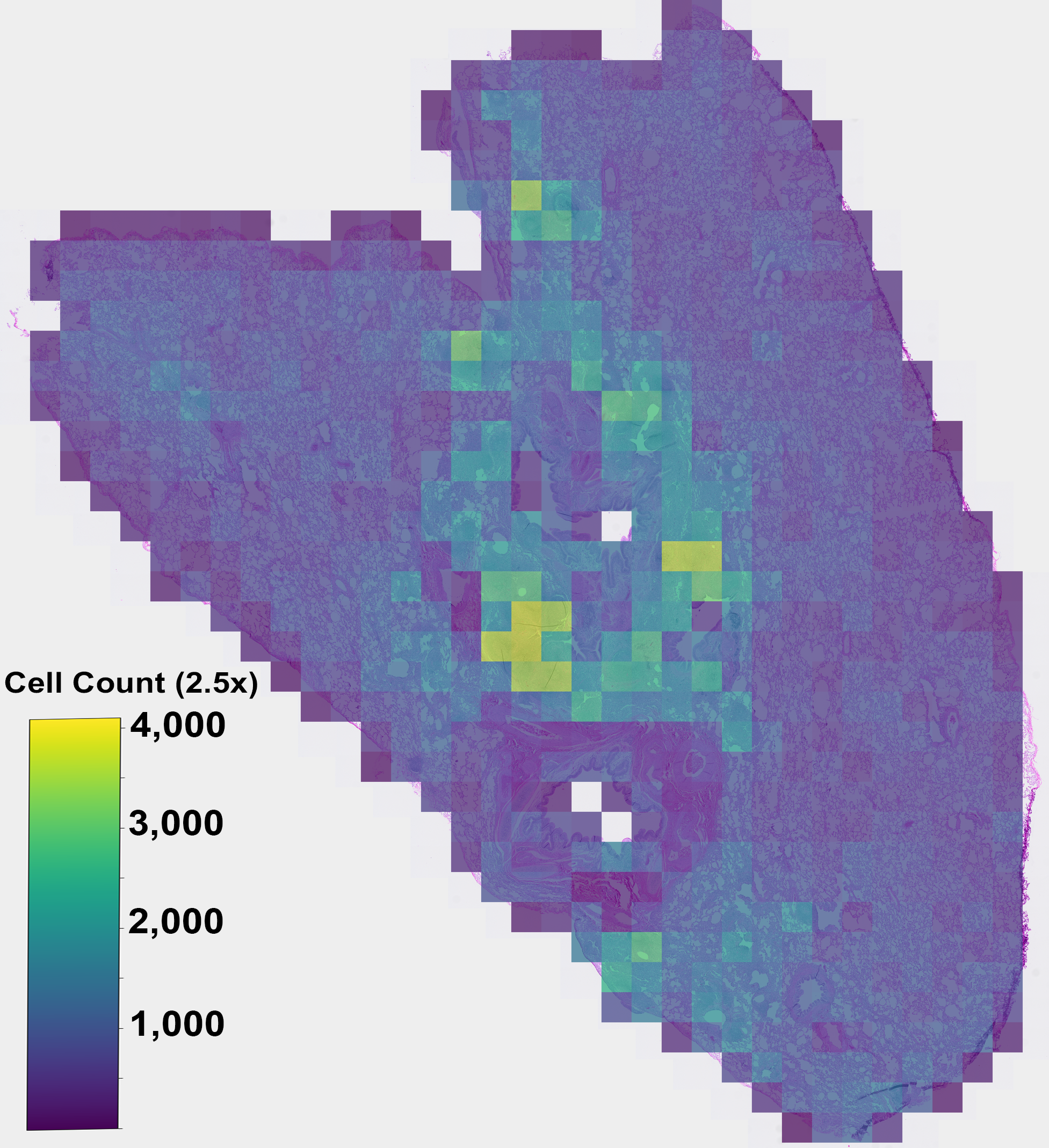}\label{fig:heatmap_10x_B}}
\caption{\textbf{Cellular density estimation.} The predicted density map (c) closely aligns with the aggregated ground truth (b), validating the ability to quantify cellularity from low-magnification embeddings.}
\label{fig:heatmaps}
\end{figure*}

\textbf{Lung Tissue Cellular Density Estimation}
The regression analysis provided compelling evidence that the Vision Foundry pipeline enables precise quantitative biomarker extraction directly from frozen embeddings. As detailed in Table \ref{tab:cell_counts}, the model fine-tuned via DINO-MX achieved competitive performance with existing models despite being trained on only 1.25 million pathology images, and significantly outperformed the generic DINOv2-giant model. This quantitative success is mirrored qualitatively in Figure \ref{fig:heatmaps}, where the predicted density maps faithfully reconstruct the complex spatial gradients, effectively distinguishing between hyper-cellular tumor nests and hypo-cellular alveolar spaces. These results confirm that the self-supervised embeddings inherently capture the structural proxies of cellularity, allowing a simple linear layer to decode this information with high fidelity. Crucially, this capability empowers users to achieve state-of-the-art results on unique datasets that are unrepresented in existing foundational models.

\begin{table}[H]
\centering
\caption{\textit{Cell Count Regression Performance. Comparison of the fine-tuned DINO-MX model against foundational baselines for downstream cellularity estimation.}}
\label{tab:cell_counts}
\setlength{\tabcolsep}{2pt}
\small
\begin{tabular*}{\textwidth}{@{\extracolsep{\fill}}lccccc}
\toprule
\textbf{Metric} & \textbf{DINO-MX} & \textbf{DINOv2-giant} & \textbf{Prov-GigaPath} & \textbf{UNI} & \textbf{Virchow2} \\
 & \textit{(Fine-tuned)} & & & & \\
\midrule
\textbf{Pearson ($r$)} & 0.9730 & 0.9519 & 0.9803 & 0.9728 & \textbf{0.9839} \\
\textbf{$R^2$}         & 0.9429 & 0.9013 & 0.9609 & 0.9443 & \textbf{0.9679} \\
\textbf{MAE}           & 64.69  & 71.06 & 52.32  & 56.11  & \textbf{45.87} \\
\bottomrule
\end{tabular*}
\end{table}

Beyond predictive accuracy, this experiment highlights a massive leap in practical efficiency. Traditional instance segmentation methods like Cellpose are computationally prohibitive for large-scale WSI analysis, requiring approximately 32 minutes to process a single slide at $10\times$ magnification (using an NVIDIA A6000). In contrast, our approach, regressing counts from low-magnification ($2.5\times$) embeddings, completed the same analysis in just 30 seconds, achieving a $\sim 60\times$ speedup. This dramatic speedup enables high-throughput analysis of cellularity patterns across entire patient cohorts without extensive hardware overhead.

\textbf{Coronary Calcium Risk Detection and Classification}
The evaluation on the external Stanford dataset provided a rigorous stress test for domain generalization. As shown in Table \ref{tab:nongated_comparison}, our model (CARD-ViT), which was trained \textit{only} on high-quality gated scans, achieved an accuracy of \textbf{0.707} and a Cohen’s $\kappa$ of \textbf{0.528} on noisy non-gated scans.

\begin{table*}[htbp]
\centering
\caption{Cross-domain performance on the Stanford non-gated test set. CARD-ViT (trained on gated data) demonstrates strong generalization when compared to AI-CAC \cite{nejm_model} (trained on non-gated data).}
\label{tab:nongated_comparison}
\small
\begin{tabular}{l|ccccc|ccccc}
\toprule
\multirow{2}{*}{\textbf{CAC Range}} & \multicolumn{5}{c|}{\textbf{CARD-ViT (Gated $\rightarrow$ Non-Gated)}} & \multicolumn{5}{c}{\textbf{AI-CAC (Trained on Non-Gated)}} \\
\cmidrule(lr){2-6} \cmidrule(lr){7-11}
& Sens & Spec & PPV & NPV & F1 & Sens & Spec & PPV & NPV & F1 \\
\midrule
0--10  & 0.933 & 0.720 & 0.778 & 0.911 & 0.848 & 0.876 & 0.770 & 0.800 & 0.856 & 0.836 \\
11--100 & 0.415 & 0.896 & 0.500 & 0.860 & 0.453 & 0.463 & 0.890 & 0.514 & 0.869 & 0.487 \\
101--400 & 0.438 & 0.942 & 0.583 & 0.901 & 0.500 & 0.406 & 0.936 & 0.542 & 0.895 & 0.464 \\
400+ & 0.593 & 0.972 & 0.762 & 0.940 & 0.667 & 0.778 & 0.955 & 0.724 & 0.966 & 0.750 \\
\midrule
\textbf{Overall (Acc/$\kappa$)} & \multicolumn{5}{c|}{\textbf{0.707 / 0.528}} & \multicolumn{5}{c}{\textbf{0.707 / 0.542}} \\
\midrule
\midrule
\multicolumn{11}{c}{\textit{Confusion Matrices (Ground Truth $\times$ Predicted)}} \\
\midrule
\textbf{Ground Truth} & \multicolumn{5}{c|}{\textbf{CARD-ViT Predicted}} & \multicolumn{5}{c}{\textbf{AI-CAC Predicted}} \\
\cmidrule(lr){2-6} \cmidrule(lr){7-11}
& 0--10 & 11--100 & 101--400 & 400+ & & 0--10 & 11--100 & 101--400 & 400+ & \\
\midrule
0--10  & 98 & 5 & 2 & 0 & & 92 & 6 & 5 & 2 & \\
11--100  & 24 & 17 & 0 & 0 & & 21 & 19 & 1 & 0 & \\
101--400 & 2 & 11 & 14 & 5 & & 2 & 11 & 13 & 6 & \\
400+  & 2 & 1 & 8 & 16 & & 0 & 1 & 5 & 21 & \\
\bottomrule
\end{tabular}
\end{table*}

To contextualize this performance, we compared it against a baseline model (AI-CAC \cite{nejm_model}) that was trained directly on non-gated data. Remarkably, our cross-domain model matched the baseline's performance (0.707 accuracy), effectively bridging the domain gap without needing the target domain's training data. Specifically, the model demonstrated high sensitivity (0.933) in identifying patients with zero calcium (score 0-10), proving it can reliably screen out healthy individuals even in standard chest CTs. While both models struggled with intermediate risk categories due to the inherent noise in non-gated imaging, these results confirm that Vision Foundry's self-supervised training learns anatomical features robust enough to withstand significant changes in imaging protocols and motion artifacts.

\section{Discussion}

The primary contribution of Vision Foundry is bridging the gap between the theoretical promise of self-supervised learning and its practical application in medicine. By abstracting the complexities of distributed training and infrastructure management, the platform eliminates the steep technical debt that has historically hindered domain experts. Our results validate that the modular strategies integrated into the DINO-MX framework, such as Magnification-Aware Distillation and domain-specific augmentation, are essential for capturing clinical semantics. This was evidenced by the superior segmentation performance in neuropathology and the robust domain generalization observed in coronary calcium scoring, confirming that medical-specific objectives yield representations significantly better than generic baselines.

Furthermore, the system demonstrates the computational efficiency of the shared-compute paradigm, where a single frozen foundational backbone supports diverse downstream tasks. This approach avoids the prohibitive cost of end-to-end fine-tuning for gigapixel images and accelerates inference speeds by orders of magnitude compared to traditional segmentation methods. Despite these advantages, the current iteration of the platform relies on high-performance computing resources, such as H100 GPU clusters, which may limit accessibility in resource-constrained clinical settings. Additionally, the validation was restricted to unimodal imaging tasks, leaving the integration of multi-modal data sources like genomics and electronic health records as a necessary direction for future development.

\section{Conclusion}

In conclusion, Vision Foundry provides a secure and reproducible environment that democratizes access to state-of-the-art computer vision. By combining the flexibility of DINO-MX with a code-free interface and HIPAA-compliant data governance, the platform empowers researchers to build, adapt, and deploy foundational models autonomously. The strong performance across diverse modalities confirms the robustness of the system, laying the groundwork for future expansions that will address current limitations by incorporating multi-modal learning to synthesize imaging features with broader clinical metadata.

\subparagraph{Acknowledgments}
This research was supported in part by the National Institutes of Health under award number UL1TR001998. The content is solely the responsibility of the authors and does not necessarily represent the official views of the NIH.

\makeatletter
\renewcommand{\@biblabel}[1]{\hfill #1.}
\makeatother

\bibliographystyle{vancouver}
\bibliography{amia}  

\end{document}